\documentclass[aps,prl,twocolumn,showpacs,preprintnumbers]{revtex4-1}

\usepackage{psfrag,graphicx}
\usepackage{dcolumn}
\usepackage{amsmath,amssymb}
\usepackage{bm}
\usepackage{amsfonts,amssymb,amsmath}        
\usepackage{epstopdf}
\usepackage{amsthm}

\newcommand{\beq}{\begin{equation}}
\newcommand{\eeq}{\end{equation}}
\newcommand{\bea}{\begin{eqnarray}}
\newcommand{\eea}{\end{eqnarray}}

\newcommand{\ket}[1]{\left | \, #1 \right\rangle}

\begin{document}

\title{A family of stabilizer codes for $D({\mathbb Z}_2)$ anyons and Majorana modes}
\author{James R.~Wootton}
\affiliation{Department of Physics, University of Basel, Klingelbergstrasse 82, CH-4056 Basel, Switzerland}

\date{\today}

\begin{abstract}

We study and generalize the class of qubit topological stabilizer codes that arise in the Abelian phase of the honeycomb lattice model. The resulting family of codes, which we call `matching codes' realize the same anyon model as the surface codes, and so may be similarly used in proposals for quantum computation. We show that these codes are particularly well suited to engineering twist defects that behave as Majorana modes. A proof of principle system that demonstrates the braiding properties of the Majoranas is discussed that requires only three qubits.

\end{abstract}

\maketitle

\section{Introduction}

Quantum error correction is an important aspect of fault-tolerant quantum computation, and many quantum error correcting codes have been proposed to serve this purpose. Recently, a great deal of interest has been focussed on so-called topological codes, such as the surface codes \cite{dennis} and quantum double models \cite{double}. These support exotic quasiparticles known as anyons, which can be used to implement fault-tolerant quantum gates on encoded information \cite{pachos_book,wootton_rev}.

The anyon model supported by a given code is classified as either Abelian or non-Abelian. This distinction determines a great deal about how the anyons behave, and how they may be used. However this is not always so clear cut, since methods can be applied to Abelian codes that give them properties that resemble non-Abelian ones \cite{wootton_ising,wootton_z6}. The most prominent method is engineering so-called twists in the code \cite{bombin_twist}. These twists are related to symmetries of the model, and so are symmetry defects \cite{you:12}.

A well-known model that is capable of supporting both Abelian and non-Abelian phases is the honeycomb lattice model \cite{honey}. This is not strictly an error correcting code, but is instead an interacting spin Hamiltonian for which anyons emerge as excitations. However, it is known that Abelian codes can be derived from the Abelian phase of this model. It is such codes that we study and generalize here. We show that these codes are well-suited to engineering twists that behave as Majorana modes. In fact they are so well-suited that we need not explicitly use the concept of twists, and can reinterpret the codes in terms of Majoranas.

\section{Matching Codes}

The Abelian phase of the honeycomb lattice model is typically studied by a perturbative analysis, with the surface code emerging in the low energy subspace \cite{honey}. However, a topological code also emerges  when the full spectrum is taken into account. It is this approach that we will take here.

Though the codes we consider are inspired by the interacting spin Hamiltonian of the honeycomb lattice model, we will not study them in this context. No Hamiltonian will be considered to energetically suppress errors. Instead will use the standard framework of quantum error correction, in which errors are detected and suppressed by continuous syndrome measurements only. This frees us from some of the restrictions required for realistic Hamiltonians, and allows us to generalize to a large family of codes. We will refer to these, which all take the form of qubit stabilizer codes, as `matching codes'.

For the purpose of presentation, it is advantageous to start with the general form of these codes. The specific cases relevant to the honeycomb lattice model will then be discussed later.

Matching codes are defined on trivalent lattices with a qubit on each vertex. For simplicity we will restrict to lattices with periodic boundary conditions. Each edge (or link) of the lattice is labelled $x$, $y$ or $z$ such that no links of the same type are adjacent. Examples are shown in Fig. \ref{lattices}. For each link, $l$, of type $\alpha \in \{x,y,z\}$ one defines a link operator $K_l = \sigma^\alpha_j \sigma^\alpha_k$ that acts on the two adjacent qubits. These form a basic set of operators used to construct the stabilizer generators of the codes.

\subsection{Stabilizer Generators}

Many different types of stabilizer code can be defined on spin lattices. To restrict to a certain class, we consider only stabilizer generators that are products of link operators. Since this still leaves considerable freedom, we will also restrict to products for which the links involved form a path.

A path is a subset of the links of the lattice such that no more than two vertices are incident upon an odd number of path edges. If there are no such vertices, we call the path a loop. If there are two then the path is a string with these vertices as endpoints. We will denote a path with endpoints $j$ and $k$ as $P(j,k)$.

\begin{figure}[t]
\begin{center}
{\includegraphics[width=8cm]{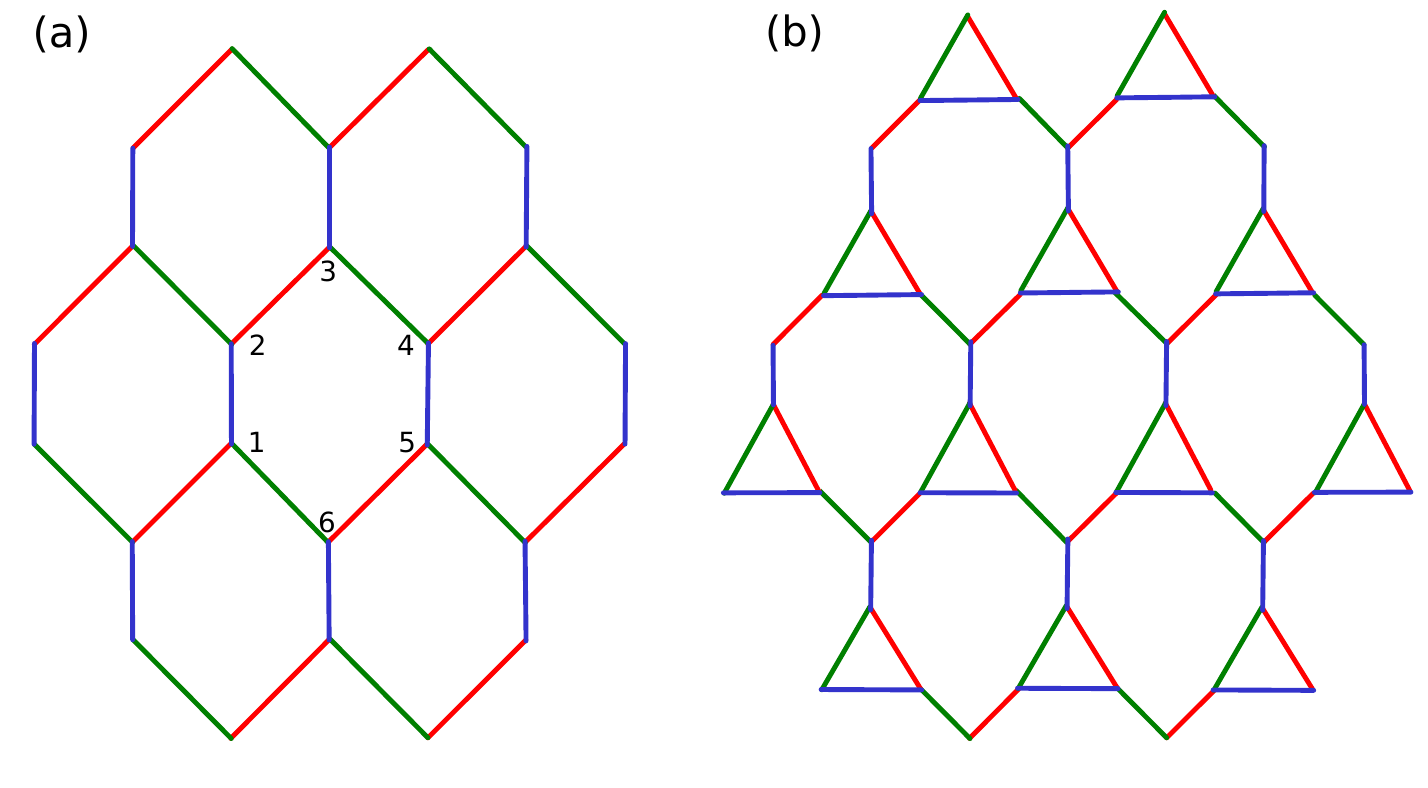}}
\caption{\label{lattices} We consider the two lattices shown, each wrapped around a torus. (a) $x$ links are right leaning diagonal lines, $y$ links are left leaning and $z$ links are vertical. These are shown in red, green and blue, respectively. Numbering of qubits around a plaquette is shown. (b) Modified honeycomb lattice. Vertices on one sublattice are replaced by triangles. The labeling of the new edges is uniquely defined by the condition that no two links of the same type are adjacent.}
\end{center}
\end{figure}

With the lattices we consider, the simplest form of loop are those that encircle each plaquette. For a plaquette $p$, we then define the following plaquette operator using this loop
\beq
W_p \sim \prod_{l \in p} K_l.
\eeq
This product will yield a tensor product of Pauli operators with a phase of $\pm 1$ or $\pm i$. The $\sim$ signifies that only the former is taken to be $W_p$, with the phase either ignored or chosen according to convenience. For the honeycomb lattice, shown in Fig. \ref{lattices}(a), the plaquette operators are
\beq
W_p = \sigma^x_1 \sigma^y_2 \sigma^z_3 \sigma^x_4 \sigma^y_5 \sigma^z_6.
\eeq
Each plaquette operator commutes with all link operators and products thereof. The plaquette operators therefore mutually commute. Also, note that any loop operator can be expressed as product of the plaquette operators.

We will consider codes for which the plaquette operators are included within the stabilizer. For a lattice of $N$ vertices (and therefore qubits) there are $N/2$ plaquettes. At least a further $N/2$ are required in order to uniquely specify a stabilizer state.

To define the additional stabilizers we consider a matching of the vertices, $M$. This is simply a set of disjoint pairs of vertices. For each pair $(j,k)$ a path $P(j,k)$ is chosen such that the vertices $j$ and $k$ form the endpoints. The following stabilizers are then defined
\beq
S_{j,k} \sim \prod_{l \in P(j,k)} K_l.
\eeq
Again the $\sim$ denotes that the phase is ignored and only the resulting tensor product of Pauli operators is used for the $S_{j,k}$.

The string stabilizers anticommute with all link operators incident upon one (but not both) of their endpoints, and commute otherwise. The same is true for products of link operators along a string: they anticommute only if they share a single endpoint and commute otherwise. Since the pairs used to define the stabilizers are non-overlapping, they will share no endpoints with each other. They therefore mutually commute with each other and the plaquette operators.

Clearly any pairing of $N$ vertices will yield $N/2$ pairs, and so there will be $N/2$ such stabilizers. Using both the loop stabilizers $W_p$ and string stabilizers $S_{j,k}$ we have a set of $N$ stabilizer generators.

\subsection{Properties of the Stabilizer}

With a set of stabilizer generators defined, we can now determine the properties of the resulting stabilizer code. Given all the chosen strings $P(j,k)$ we can determine $n_l$, the number of strings each link, $l$, is included in. We refer to links as `even' or `odd' depending on whether their $n_l$ is even or odd. As such, clearly
\beq \label{linkprod}
\prod_{(j,k) \in M} S_{(j,k)} \sim \prod_{l \in \rm{odd}} K_l
\eeq
since any even power of the link operators yields the identity.

For each vertex $j$ we can also consider the quantity $\nu_j = \sum_{l \in j} n_l$, where the sum is over the three links incident upon $j$. Since all vertices are the endpoint of a single string, the quantity $\nu_j$ will be odd for all $j$. As such, an odd number of odd links must be incident upon each vertex. Since all vertices are trivalent, this implies that there will also be an even number of even links incident upon each vertex.

Due to this property, the set of even links corresponds to a path (or set of paths) that form a loop (or loops). The plaquettes may then be bicoloured according whether or not they are enclosed by the set of loops. These colours are referred to as `black' and `white', and the corresponding sets of plaquettes are denoted $b$ and $w$.

Two plaquettes are neighbouring if their boundaries share at least one link. If any of these is an odd link, the two plaquettes will clearly be enclosed by the same loop of even links. They will therefore be of the same colour. Similarly, two neighbouring plaquettes that do not share any odd links will be separated by a loop of even links, and so will be different colours.

Given this bicolouring, we can find similar relations to Eq. (\ref{linkprod}) for the plaquettes 
\beq \nonumber
\prod_{p \in b} W_p  \sim \prod_{p \in w} W_p  \sim \prod_{l \in even} K_l .
\eeq
It is straightforward to see that $ \prod_{l} K_l \sim \openone$, and so the product over even links above can be substituted with a product over odd ones. Combining with Eq. (\ref{linkprod}) we then find the following relation between the three types of stabilizer
\beq
\prod_{p \in b} W_p  \sim \prod_{p \in w} W_p \sim \prod_{(j,k) \in M} S_{(j,k)}.
\eeq
As such, only $N-2$ of the $N$ stabilizer generators are independent. The stabilizers may therefore be used to define a stabilizer code that can store two logical qubits. However, this is not the approach we will take.

Note that the choice of the paths $P(j,k)$ used to define the stabilizer generators does not alter the stabilizer as a whole. The product of an $S_{j,k}$ defined using a path $P(j,k)$ and an $S'_{j,k}$ defined using a path $P'(j,k)$ will be the product of the $W_p$ enclosed by the resulting loop. These alternate $S'_{j,k}$ can therefore equally be regarded as stabilizers. The paths are only required in order to define the bicolouration of the lattice.

\subsection{Examples of Matching Codes}

\begin{figure}[t]
\begin{center}
{\includegraphics[width=8cm]{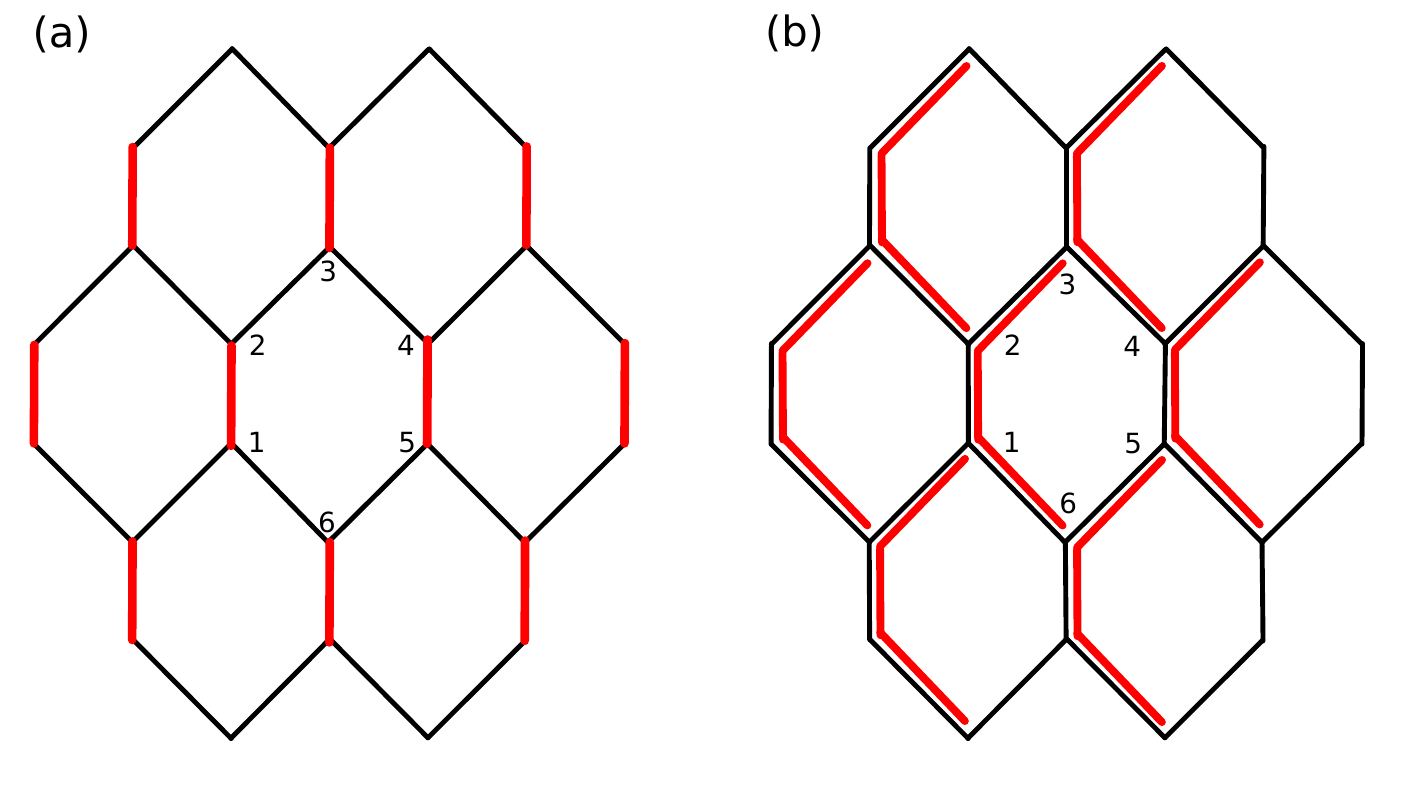}}
\caption{\label{examples} Thick red lines on the hexagonal lattice denote paths on which stabilizers are defined. (a) String stabilizers are defined on $z$-links. (b) String stabilizers correspond to the left half of each plaquette.}
\end{center}
\end{figure}

A simple subclass of matching codes are those for which only nearest neighbour pairing are allowed, and the path used is always the single link between the vertices. The case of $z$-links used as string stabilizers on the honeycomb lattice is shown in in Fig. \ref{examples} (a). Such models are well known from the Abelian phase of the honeycomb lattice model \cite{honey}. They are known to support the $D({\mathbb Z}_2)$ anyon model \cite{double}, and map to the standard square lattice toric code when violations of string stabilizers are prohibited \cite{honey}. All other such models behave similarly, though they map to toric codes on different lattices in general.

Another example of a matching code is one for which the top and bottom qubits of each plaquette are paired, with the path chosen to be the three links on the left side of the plaquette. This is shown in Fig. \ref{examples} (b). In this case, each string stabilizer is associated with a plaquette, and so can be interpreted as a second plaquette operator
\beq
S_p = \sigma^y_6 \sigma^x_1 \sigma^y_2 \sigma^x_3.
\eeq
Rather than considering the usual $W_p$ operators, we can use the following operators as stabilizer generators
\beq
S'_p = S_p W_p = \sigma^y_3 \sigma^x_4 \sigma^y_5 \sigma^x_6.
\eeq
This effectively splits each hexagonal plaquette into two square plaquettes, each with its own plaquette operator \cite{wootton_ising}. The resulting code corresponds to the Wen plaquette model \cite{wen_plaquette}, which is equivalent by local unitaries to the standard square lattice surface code \cite{brown:11}.

\subsection{Planar Boundary Conditions}

Periodic boundary conditions are assumed in the above for the simplicity of translational invariance. Practically, however, planar boundary conditions are more appropriate \cite{dennis}. These are also possible for matching codes.

As an example, consider a Wen plaquette model with planar boundary conditions. On the boundary it is difficult to apply the above interpretation of the model as a matching code. However, in the bulk the interpretation can be easily applied. The matching, $M$, for the vertices of the bulk is well defined. The specific matching of the Wen plaquette model is not required for consistency of the bulk and boundary, and so any $M$ may be used. We are therefore free to choose any matching code for the bulk, with an assurance that this will be consistent with the standard boundary.

\section{Anyon Model}

The stabilizer states of the matching codes can be interpreted in terms of anyonic quasiparticles. An anyon is said to reside on a plaquette $p$ for a state $\ket{\psi}$ if $W_p\ket{\psi} = -\ket{\psi}$, and similarly on a string $P(j,k)$ if $S_{j,k} \ket{\psi} = -\ket{\psi}$. An eigenvalue of  $+1$ is interpreted as anyonic vacuum in both cases.

In the examples of matching codes above, the anyon model corresponds to $D({\mathbb Z}_2)$. This is true of all matching codes, as we will now show. To do this we consider stabilizer states, which correspond to states of definite anyon configurations. The effect of applying Pauli rotations to the stabilizer states is to map between them, corresponding to the creation, transport and annihilation of anyons. Using these operators we can therefore determine the braiding and fusion properties of the anyon model.

\subsection{Anyons on strings}

First let us consider the application of link operators. For a link $l=(j,j')$ the link operator $K_l$ will commute with all stabilizers if $(j,j')$ is a pairing of $M$. Otherwise it will anticommute with the two string stabilizers $S_{j,k}$ and $S_{j',k'}$.

Applying the link operator to a state with definite eigenvalues for these stabilizers will have the effect of flipping the eigenvalue from $+1$ to $-1$, or vice-versa. The corresponds to the creation (annihilation) of an anyon pair if the two strings initially held vacuum (anyons). If only one string initially held an anyon, the effect is to move it to the other. 

Consider a graph for which all the pairs of vertices in $M$ are combined to form single vertices, with each new vertex inheriting the edges from both its predecessors. The anyons created by link operators can be thought of as residing on the vertices of this graph. The anyons can be moved between pairs of vertices connected by an edge. Since the original lattice had no disjoint subgraphs, this property is inherited by the new lattice. It is therefore possible for an anyon to move from any string stabilizer to any other. As such, the same species of anyon must live on all string stabilizers. Let us use $\epsilon$ to denote this anyon type. Since it is created in pairs, it is clear that it obeys the fusion rule $\epsilon \times \epsilon = 1$.

Consider three vertices, $i$, $j$ and $k$, where both $i$ and $k$ are neighbours of $j$. We choose this triplet such that each is the endpoint of a unique string operator, which is always possible in general. For $j$ the other endpoint is at vertex $j'$, which has a neighbour $j''$. Let us start with a state in which both the strings of $i$ and $k$ hold an $\epsilon$, whereas that of $j$ does not. Applying the link operator $K_{i,j}$ will move the $\epsilon$ on the string of $i$ to that of $j$. Applying $K_{j',j''}$ then moves it again to the string of $j''$. Applying $K_{j,k}$ moves the $\epsilon$ on the string of $k$ to that of $j$, and $K_{i,j}$ moves this on to the string of $i$. Finally, $K_{j',j''}$ followed by $K_{j,k}$ moves the $\epsilon$ on the string of $j''$ to that of $k$. The final effect is an exchange of the two $\epsilon$ anyons. The total operator used to realize this is
\beq \nonumber
 K_{j,k} K_{j'j''} K_{i,j} K_{j,k} K_{j',j''} K_{i,j} = K_{j,k} K_{i,j} K_{j,k}  K_{i,j} = - \openone.
\eeq
This demonstrates that exchange leads the wave function to acquire a phase of $-1$. The $\epsilon$ anyons are therefore fermions.

\subsection{Anyons on plaquettes}

To determine the anyon types living on plaquettes, consider a link $l$ shared by two neighbouring plaquettes $p$ and $p'$. We use $\alpha \in \{x,y,z\}$ to denote its type and $j$ to denote either of the vertices that it is incident upon. The application of $\sigma^\alpha_j$ to the qubit on vertex $j$ will anticommute with the plaquette operators of $p$ and $p'$ and no others. It therefore corresponds to the creation of a pair of plaquette anyons. However, we must consider also the additional effects on the string stabilizers.

If $l$ is an odd link, it follows that the other two links incident upon $j$ will either be both odd or both even. The support of the product of all string stabilizers on $j$ will therefore be $\openone$ or $\sigma^\alpha$, respectively. In either case this product will commute with the applied operation $\sigma^\alpha_j$. It therefore results in the creation of an even number of $\epsilon$ anyons.

If $l$ is an even link, it follows that exactly one of the other two links incident upon $j$ will be odd. This results in a product of all string stabilizers that anticommutes with $\sigma^\alpha_j$. The application of this rotation therefore creates an odd number of $\epsilon$ anyons.

From these properties we see that creating a pair of anyons on neighbouring plaquettes of the same colour (i.e., separated by an odd link) will result also in the creation of an even number of $\epsilon$ anyons. These may then be removed by the application of link operators. An operation that creates a pair of plaquette anyons same coloured neighbouring plaquettes with no additional $\epsilon$ anyons will therefore always exist. This clearly holds also for non-neighbouring plaquettes of the same colour. All anyons on same coloured plaquettes therefore belong to the same anyon type.

For different coloured plaquettes (separated by an even link) there will always be an odd number of $\epsilon$ anyons created. The application of link operators can reduce these to a single $\epsilon$, but this cannot be removed. Anyons that live on different coloured plaquettes must therefore belong to different species, which differ up to fusion with an $\epsilon$. We use $e$ to denote the anyons on white plaquettes and $m$ for those on black. These properties can be summarized in the fusion rules
\beq
e \times e = m \times m = \epsilon \times \epsilon = 1, \,\,\, e \times m  = \epsilon.
\eeq
These rules completely generate the fusion rules of the model. These are the fusion rules of the $D({\mathbb Z}_2)$ anyon model. This has three non-trivial anyon types. One is a fermion, which we have already shown to be $\epsilon$. The two others have bosonic exchange properties with respect to themselves, and semionic with respect to each other.

\section{Majorana Mode Interpretation}

Studies of the honeycomb lattice model often make use of a mapping of the problem from qubits to Majorana modes \cite{honey}. Specifically, the qubit operators are mapped to Majorana operators by associating four Majorana modes to each vertex. Three of these modes for each vertex are absorbed into a lattice gauge theory. With the remaining Majorana mode $c_j$ at each vertex $j$, the link operators can be expressed,
\beq
K_l = \left(i c_j c_k \right)  u_{jk} 
\eeq
Here the $u_{jk}$ term comes from the lattice gauge theory. The $i c_j c_k$ term is the parity operator for the Dirac mode that is composed of the two Majorana modes $c_j$ and $c_k$.

Due to the property $c_j^2=1$ for Majorana modes, the string stabilizers may be expressed
\beq
S_{j,k} \sim \left(i c_j c_k \right) \prod_{l \in P(j,k)} u_{jk}.
\eeq
This is the parity operator for the Dirac mode that consists of the Majoranas at the endpoints of the string $P(j,k)$, again with a factor from the gauge theory. The plaquette operators consist only of gauge theory terms with no Majoranas.

This form for the stabilizers agrees exactly with what we know about their anyonic occupations. The string stabilizers are parity operators for a Dirac mode, since their eigenvalue $\pm 1$ signals the presence or absence of a fermion.

The path dependent factor comes from the fact that these Dirac fermions can decay into an $e$ and an $m$. The occupation of their Dirac modes will therefore depend on the bicolouring of the lattice, which depends on the paths taken by the string stabilizers.

If we consider only states for which there is always vacuum on the plaquettes, the path dependence of the string stabilizers is effectively removed. The system can then simply be interpreted as one for which a single Majorana mode is pinned to each vertex. The string stabilizers simply correspond to an arbitrarily chosen pairing of the Majorana modes, and are defined as the parity operator for the Dirac mode of the pair.

The stabilizer states are those for which there is a definite pairing of all the Majorana modes, and each of the corresponding Dirac modes have a definite occupation. For these states, the Majorana interpretation is nothing more than a mathematical curiosity. However, as we shall see in the next section, it is possible to use these Majoranas to store quantum information, and to braid them in order to process this information. Their full nature as non-Abelian anyons can therefore be realized. As such the matching codes, which would normally be interpreted as Abelian models, may also be interpreted and used as non-Abelian anyon models.

Note that this interpretation is related to the concept of `twists', which behave as Majorana modes \cite{honey,bombin_twist}. These have been considered in many cases including surface codes \cite{bombin_twist,you:12,brown:13} and the honeycomb lattice model \cite{petrova:14}. These can be used for quantum computation, and have better resource usage than other surface code based approaches \cite{hastings:14}.

The study of twists usually considers Majorana modes as being associated with lattice dislocations. The Majoranas are created and annihilated by deforming the lattice on which the code is defined. The Majorana modes of the matching codes, on the other hand, do not require such a complex interpretation. They are always present and do not require lattice deformations. This makes the matching codes a useful framework in which to study these defects.

\section{Braiding of Majorana Modes}

Let us consider only the anyonic vacuum state (i.e. the stabilizer space) for all stabilizers. In order to add some degeneracy into the stabilizer space, we remove some of the link stabilizer operators from the stabilizer. The corresponding Majoranas are then unpaired. Parity operators can be defined using pairings of these. Different pairings correspond to different bases in which the corresponding stabilizer space states can be measured. We refer to these unpaired Majoranas as computational Majoranas, and the rest as background pairs.

\begin{figure}[t]
\begin{center}
{\includegraphics[width=8cm]{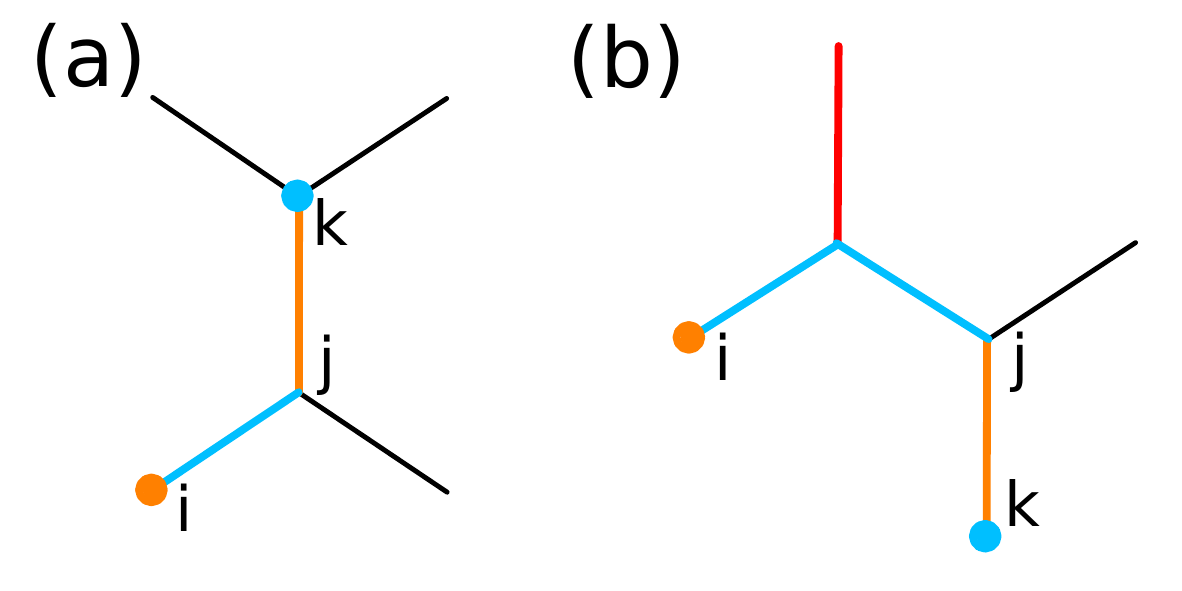}}
\caption{\label{teleport} Two processes by which a computational Majorana on vertex $i$ is moved to vertex $k$. The path between $j$ and $k$ (orange) corresponds to a stabilizer of the initial state in both cases. The path between $i$ and $j$ (blue) corresponds to a stabilizer of the final state. In (b) the additional vertical line (red) is a stabilizer for both the initial and final states.}
\end{center}
\end{figure}

In order to achieve some degree of quantum computation with the computational Majoranas, we must be able to braid them with each other. Since all Majoranas are pinned to vertices, we cannot move them freely. However, we can hope to perform exchange operations that result in their effective movement around the lattice. This can be done using so-called `anyonic state teleportation' \cite{bonderson:08}.

The simplest possible operation is to exchange a computational Majorana, $c_j$, with a neighbouring background pair. Two examples of this are shown for the honeycomb lattice in Fig. \ref{teleport}.

The process shown in Fig. \ref{teleport} (a) exchanges the computational Majorana on vertex $i$ with the background pair in the link $(j,k)$, resulting in the computational Majorana moving to vertex $k$. To do this we measure the link operator $K_{i,j}$, which will result in a random outcome. The resulting state will be an eigenstate of $K_{i,j}$, and will no longer be stabilized by $S_{j,k}$. The set of pairs, $M$, is therefore altered by replacing the pair $(j,k)$ with $(i,j)$. If the outcome of the measurement is $+1$, the resulting state holds vacuum for the new stabilizer $S_{i,j} = K_{i,j}$ as well as all previous stabilizers. This process corresponds to the background pair on $(j,k)$ moving to $(i,j)$, and the computational Majorana on $i$ moving to the other side of this background pair. All other Majoranas remain stationary. For the outcome $-1$ the process has the same effect, except that an $\epsilon$ is fused with both the moved background pair and computational Majorana. To undo the effect of this, the link operator $K_{j,k}$ is applied. The state will then be the same as if the outcome was $+1$.

For a bipartite lattice, the above process will only allow computational Majoranas to move around the same sublattice. In order for them to move between sublattices, exchanges such as in Fig. \ref{teleport} (b) must be applied. The only difference between this and the above is that the measured operator $S_{i,j}$ is a product of two link operators.

The exchange of a Majorana with a background pair is a trivial operation, since the net sector of a such a pair is vacuum. Nevertheless the operation would be expected to yield a global phase that depends on the chirality of the exchange. However it is evident that this phase, and the distinction between clockwise and anti clockwise exchanges around a background pair, does not arise in the process described above. As such the braid statistics realized by the Majoranas are projective braid statistics, which are the same as the full braid statistics up to global phases \cite{barkeshli:13}.

With the ability move the computational Majoranas around, we can consider the effects of their braiding. Since this involves moving background pairs out of the way, different paths taken by Majoranas between two points will result in different configurations of the background pairs. In order to avoid any ambiguity this may cause when assessing the effects of the exchange, we will consider only braid operations for which the initial and final states have the same background configuration.

For concreteness we will use the modified honeycomb lattice. The only string stabilizers that will be considered are those for the pairing of nearest neighbours, and so are defined on a single link. The set of pairs, $M$, then becomes the set of links for which there are stabilizers. We use $S_l$ to refer to the stabilizer for the links $l \in M$. This lattice is not bipartite, and so the Majoranas may be moved between any pair of vertices by measuring single link operators only.

One possible choice of links for $M$ is simply all those connected by $z$-links, and so each $z$-link operator becomes a stabilizer. Using this as a background, we will define a code for which computational Majoranas are placed on a 1D line.

\begin{figure}[t]
\begin{center}
{\includegraphics[width=8cm]{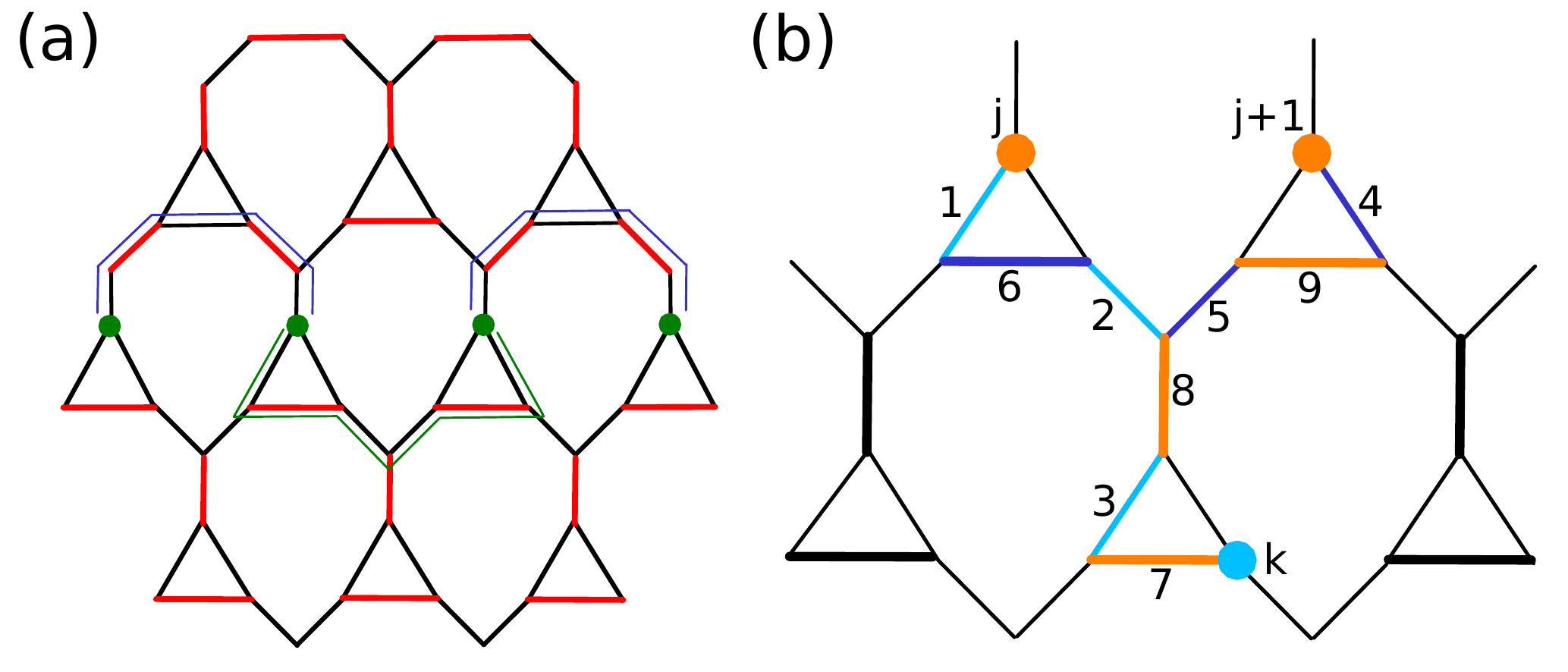}}
\caption{\label{exchange} Stabilizers on the modified honeycomb lattice that lead to a line of unpaired Majoranas. (a) Thick red links correspond to pairings within $M$. The paths $P(j,j+1)$ along which the Majoranas are created pass above the Majorana line. These are shown with thin blue lines. The path $P_j$ along which parity operators are defined goes below the Majorana line. This is shown with a thin green line. (b) Numbering of links used for the exchange operation. Those in bold are part of $M$. Different coloured links are used at different points in the exchange process. Links $1$, $2$ and $3$ are coloured light blue, $4$, $5$ and $6$ are dark blue, $7$, $8$ and $9$ are orange.}
\end{center}
\end{figure}

Consider a 1D row of vertical $z$-links for which every $d$th $z$-link is chosen to be removed from the stabilizer. We call these `flagged' links. For the $j$th flagged link from the left, we use $j$ to refer to the lower vertex. For every odd $j$ a path $P(j,j+1)$ is then found, such that the $j$th and $j+1$th flagged links the are first and last links of the path, and the path has endpoints on the vertices $j$ and $j+1$. Furthermore, the links of the paths should alternate between links that are part of the original pairing (i.e. $z$-links) and those which are not. The set of pairings $M$ is then modified by removing all $z$-links along these paths, and adding all other links along them. This results in a code for which computational Majoranas are located on all vertices $j$. An example of this for $d=1$, and so all links along the row removed, is shown in shown in Fig. \ref{exchange}(a).

For each neighbouring pair of computational Majoranas $(j,j+1)$ we require a path in order to define the corresponding parity operator. Rather than used the same paths used to generate the code above, we use an alternative one which we denote $P_j$. Such a path is shown in \ref{exchange}(a). These paths are chosen such that they alternate between links that are within $M$ and those that are not, except at a mid-point at which two links not in $M$ will be connected. The parity operator is then defined,
\beq
\pi_{j,j+1} = i \prod_{l \in P_j} K_j.
\eeq
Here the product is taken sequentially along the path. The $j$ end corresponds to the right-most factors and the $j+1$ end to the left-most.

There will always be an even number of links along the path, both for the specific example of \ref{exchange}(a) and those for which the distance between computational Majoranas is increased. This means the parity operators have support on an odd number of qubits, and so the factor of $i$ is required to ensure that their eigenvalues are $\pm 1$. It also means that
\beq
\pi_{j,j+1} = -\pi_{j+1,j},
\eeq
where the latter is the same product of link operators but in reserved order. This reflects the $i c_j c_{j+1} = - i c_{j+1} c_j$ behaviour expected for the parity operators when the order of the Majorana operators is reversed. The commutation relations for parity operators with each other are also what we expect: those for disjoint pairs of Majoranas will commute, whereas those that share a single Majorana will anticommute.

We explicitly consider the exchange of the two Majoranas shown in green in Fig. \ref{exchange}(b), using $\ket{\psi}$ to denote the initial state. To perform a anticlockwise exchange we first move the Majorana at $j$ away from its initial position. Specifically, we move it diagonally downwards and towards the right to the vertex $k$. This is done by sequential applications of the anyonic teleportation procedure on the light blue edges. The resulting state is
\beq
\ket{\psi'} \sim \prod_{l \in b} (\openone + K_l) \ket{\psi}.
\eeq
Here $b$ denotes the set of light blue links. The factors $(\openone + K_l)$ are projectors onto subspaces stabilized by the link operators for these edges. The product is again taken sequentially as the Majorana is moved.

Next we move the Majorana at $j+1$ to $j$. This is done by moving it diagonally downwards and to the left until it intersects with the previous path. It is then moved up that path until it reaches $j$. The teleportation for this case is achieved using the dark blue link operators. The resulting state is
\beq
\ket{\psi''} \sim \prod_{l \in B} (\openone + K_l) \ket{\psi'}
\eeq
where $B$ denotes the set of dark blue links. Finally the Majorana at $k$ is moved to $j+1$ using the orange links. The final state is
\beq
U \ket{\psi} \sim \prod_{l \in O} (\openone + K_l) \prod_{l \in B} (\openone + K_l) \prod_{l \in b} (\openone + K_l)\ket{\psi'}
\eeq
Now we must determine the effective unitary $U$ that is implemented by this exchange.

Note that the initial and final states are stabilized by the same set of link operators, $M$, which corresponds to the $z$-links. Also note that $O \subset M$. The above may therefore be rewritten 
\beq \nonumber
U \ket{\psi} \sim \prod_{l \in M} (\openone + K_l) \prod_{l \in B} (\openone + K_l) \prod_{l \in b} (\openone + K_l)\prod_{l \in M} (\openone + K_l) \ket{\psi'}.
\eeq
The factor $\prod_{l \in B} (\openone + K_l) \prod_{l \in b} (\openone + K_l)$ will yield a sum of products of the blue link operators. Some of the terms in this will anticommute with the link operators of $M$, whereas others will commute. The effects of the former will be removed by the $\prod_{l \in M} (\openone + K_l)$ factors. The latter consists of only two terms: the identity, and the product of the blue links along the path $P_j$,
\beq
\prod_{l \in b \cup P_j} K_l \prod_{l \in B \cup P_j}  K_l = - \prod_{l \in B \cup P_j}  K_l \prod_{l \in b \cup P_j} K_l.
\eeq
The order of the product over $b$ has the links at the $j$ end of $P_j$ to the left. The $B$ product has the $j+1$ end to the left. However, since both factors act on an even number of qubits, their order can be reversed without effect. The r.h.s. of the above can therefore be written as a product sequentially along $P_j$.

This term includes all of the factors of $\pi_{j,j+1}$ that are not in $M$. There will always be an even number of such factors, due to the symmetry of the path. As such, this term is related to the parity operator by
\beq
\prod_{l \in B \cup P_j}  K_l \prod_{l \in b \cup P_j} K_l = -i \pi_{j,k} \prod_{l \in M \cup P_j} K_l.
\eeq
The resulting unitary evolution of the exchange can then be written
\beq
U \ket{\psi} = (\openone + i \pi_{j,j+1}) \ket{\psi} \,\, \therefore \,\, U = \frac{\openone + i \pi_{j,j+1}}{\sqrt{2}}.
\eeq
Apart from a global phase of $e^{i\pi/8}$, this is exactly the expected result for the exchange of two Majoranas \cite{honey,pachos_book}.

For the clockwise exchange of $j$ and $j+1$, the mirror image of the above process is applied. Since the ordering of factors in each product is reversed, this will yield the effect
\beq
U^\dagger = \frac{1}{\sqrt{2}} (\openone - i \pi_{j,j+1}).
\eeq
Again, this is exactly the expected result from Majorana braiding, up to a global phase.

In order to fuse a pair of computational Majoranas, we simply need to determine the occupation of their net Dirac mode. This is done by adding their parity operator into the stabilizer and measuring it. For practical reasons, the pair should be moved close together in order for this operator to be measurable.

\subsection{Minimal Demonstration of Braiding}

\begin{figure}[t]
\begin{center}
{\includegraphics[width=4cm]{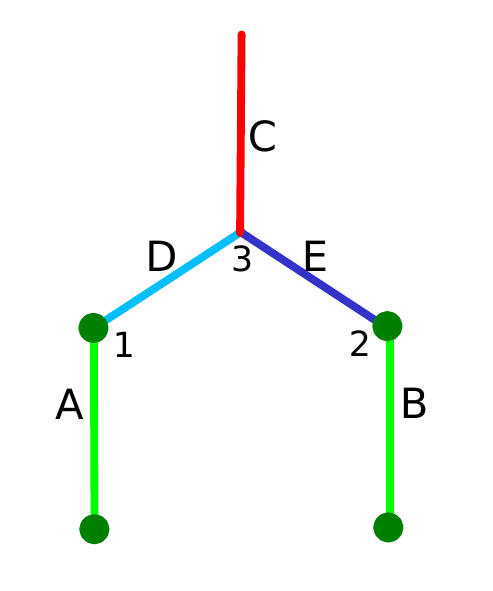}}
\caption{\label{experiment} Two pairs of computational Majoranas, with parity operators given by links $A$ and $B$. Link $C$ denotes a stabilizer.}
\end{center}
\end{figure}

The simplest example of an exchange of two computational Majoranas is shown in Fig. \ref{experiment}. The system is composed of six qubits, corresponding to three adjacent $z$-links in the honeycomb lattice and the $x$ and $y$ links that connect them. The only stabilizer is $S = K_C$. The link operators $\pi_A = K_A$ and $\pi_B = K_B$ are taken to be parity operators for their respective pairs. The fusion space of the computational anyons is four-dimensional. A basis $\{\ket{k_A,k_B}\}$ can be defined using the eigenstates of the parity operators, labelled by the eigenvalues $k = \pm 1$

A clockwise exchange of the computational Majorana at $1$ with that at $2$ can be achieved by first measuring $K_D$, then $K_E$, and then finally the stabilizer $S=K_C$. In a proof of principle experiment, we may simply post-select the results for each of these gives the outcome $+1$. Single and double exchanges have the following effect on the basis states for the computational Majoranas
\bea \nonumber
R: \, \ket{k_A,k_B} &\rightarrow & \frac{1}{\sqrt{2}}\left( \ket{k_A,k_B} + i \ket{-k_A,-k_B}\right) \\
R^2: \, \ket{k_A,k_B} &\rightarrow & \ket{-k_A,-k_B}
\eea
By preparing and then measuring these basis states, the effects of the braiding may then be shown.

This system may be made simpler still by noting that the unnumbered qubits contribute trivially. The $z$-link operators may then be truncated onto the numbered qubits: $S = \sigma^z_2$, $\pi_A = \sigma^z_1$ and $\pi_B = \sigma^z_2$. The basis states for the computational Majoranas are then simply the $z$ basis states, and measurement of the stabilizer is a single qubit $z$ measurement. The whole process then corresponds to preparation of $z$ basis states for three qubits, followed by two entangling two qubit measurements, followed by $z$ basis measurements. Using this process, the basic principle behind the non-Abelian braiding of the computational Majoranas may be demonstrated.

\subsection{Adiabatic Realization of Braiding}

In the above it is always assumed that the code is used as part of a circuit model quantum computation. As such, it is done without implementing the Hamiltonian associated with the stabilizer code \cite{ben_rev}. However, one could also consider the case in which this Hamiltonian is indeed used to enhance the error suppression of the code.

The Hamiltonian corresponding to a stabilizer code with stabilizer generators $S_j$ is typically
\beq
H_S = - \sum_j S_j.
\eeq
This results in a degenerate ground state space that corresponds to the stabilizer space of the code, and energetically penalizes states with non-trivial syndrome. For the matching codes, this Hamiltonian would need to consist of all plaquette and string stabilizers, which are many-body interactions in general. However, a more simple form arises in the case that all matchings correspond to nearest neigbour pairs, and all so all string stabilizers are link operators. In this case we can consider the Hamiltonian
\beq
H = - J \sum_{l\in M} K_l - h \sum_{l \notin M} K_l, \,\,\, h \ll J
\eeq
This Hamiltonian consists only of nearest neighbour two-body interactions between qubits. These correspond to link operators, with a higher coupling for the links within the matching than those not. When the effects of the latter as a perturbation on the former are studied, the many-body plaquette interactions emerge \cite{honey}. The Hamiltonian will then have the same ground state and as $H_S$ above, with a finite gap \cite{honey}. There will be some differences in the excitation spectrum, but these have only local effect. This Hamiltonian corresponds to the Abelian phase of the honeycomb lattice model, from which the basic matching codes are derived.

To create unpaired Majorana modes, implement the Hamiltonian for the relevant set of links, $M$. This was also studied previously in \cite{petrova:14}. To move the unpaired modes, one makes the corresponding alterations to the set $M$ as described in previous sections. The only difference is that this change in $M$ is no longer implemented on the physical system by measuring link operators. Instead it is done by slowly changing the Hamiltonian. To remove a link $l$ from $M$, and add another link $l'$ in its place, the coupling from $M$ can be slowly tuned from $J$ to $h$. That for $l'$ can then be raised from $h$ to $J$. If this is done slowly enough to obey the adiabatic theorem, no additional fermions will arise during the process as they do in the circuit model case.

By this means, unpaired Majorana modes can be engineered, transported and braided using the Abelian phase of the honeycomb lattice model. This is something usually associated with the non-Abelian phase, for which the Majorana modes are pinned to vortex excitations. Using this method, holonomic quantum computation can be implemented \cite{wootton_z6,landahl_holo,brun_holo} using the Majorana modes.

Note that the Majorana modes that emerge from a Hamiltonian in this way are similar to those that arise in nanowires \cite{kitaev:01}. However protection in the nanowire case is based on  fermionic-parity conservation, whereas matching codes have fully topological protection. This has similarly been achieved by a different (but also surface code based) approach in \cite{terhal:12}.

\section{Color codes from matching codes}

Matching codes realize the $D({\mathbb Z}_2)$ anyon model. However, they can also be used to construct further codes which realize more complex anyon models, through a process of embedding one code inside another. Here we will consider this for the specific example of the celebrated color code \cite{bombin_color}.

For any stabilizer code we can consider the basis formed by stabilizer states. In this basis we can associate a two-level quantum system with each stabilizer generator. The basis states for each of these correspond to its $+1$ and $-1$ eigenspaces.

Typically, when considering a stabilizer code, we require all these two-level systems to be in a definite state: the $+1$ eigenspace of the corresponding stabilizer. However, we could instead consider only imposing this restriction on some stabilizers. The others could then be used as if they were qubits themselves to define another stabilizer code. In this way we can embed codes within other codes.

This method is in some ways similar to concatenation. However, the purpose is very different. Concatenation combines many low distance codes to create a code of higher distance. For the embedding considered here we start with a single code, which could have arbitrarily high distance, such as surface or matching codes. The purpose of the embedding is simply to generate a new code, which may have more favourable properties than  the original.

\begin{figure}[t]
\begin{center}
{\includegraphics[width=8cm]{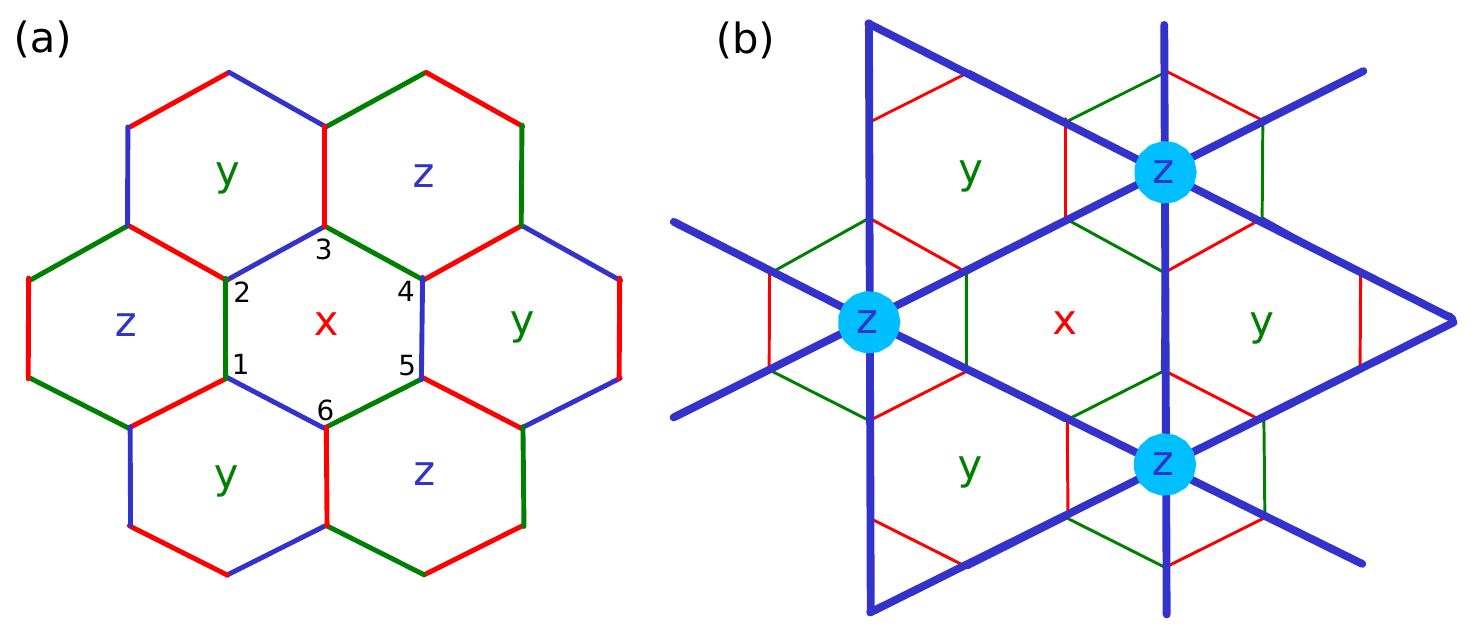}}
\caption{\label{embed} (a) Hexagonal lattice with plaquettes tricoloured $x$, $y$ and $z$. Links are tricoloured correspondingly, with $x$ and $y$ links alternating around a $z$ plaquette, etc. Again red, blue and green denote $x$, $y$ and $z$, respectively. (b) The hexagonal lattice with a triangular lattice superimposed on top. The $z$-links in the former correspond to edges in the latter, $x$ and $y$ plaquettes in the former are plaquettes in the latter, and $z$ plaquettes in the former are vertices in the latter.}
\end{center}
\end{figure}

For a specific example, let us again consider a honeycomb lattice. Rather than use the same tricolouration of links as before, we will instead use that of Fig. \ref{embed} (a). For this we define the matching code in which all $z$-links are used as string stabilizers.

With this tricolouration of links we can also consider a tricolouration of plaquettes, as shown in Fig. \ref{embed} (a). The plaquette operators then take the form
\beq
W_{p_\alpha} = \sigma^\alpha_1 \sigma^\alpha_2 \sigma^\alpha_3 \sigma^\alpha_4 \sigma^\alpha_5 \sigma^\alpha_6.
\eeq
where $p_\alpha$ is an $\alpha$-plaquette for $\alpha \in \{x,y,z\}$. 

We will use the plaquette stabilizer operators as generators in our embedded code, but not those of the $z$-links. As such each $z$-link can be thought of as a two-level quantum system in its own right, with basis states spanned by the vacuum and $\epsilon$ occupancies of the link. We will refer to these as `link qubits'.

These links qubits can be interpreted as sitting on the edges of an triangular lattice, as shown in Fig. \ref{embed} (b). The plaquettes of the original hexagonal lattice of qubits correspond to either plaquettes or vertices of the triangular lattice of link qubits. Specifically the $x$- and $y$-plaquettes of the honeycomb lattice are plaquettes of the triangular lattice and the $z$-plaquettes are vertices.

Let us now define a surface code on the triangular lattice of link qubits, such that we have a surface code embedded in a matching code. When defined on a lattice with qubits on edges, a surface code has stabilizers
\beq
B_p = \prod_{j \in v} \sigma^x_j, \,\,\, A_v = \prod_{j \in p} \sigma^z_j.
\eeq
For the lattice of links we must determine what link qubit operators will be used in place of these. This means finding substitutes of $\sigma^z$ and $\sigma^x$.

Let us associate the $\sigma^z$ basis with the $\epsilon$ occupancy basis, and so use the string operators $K_j$ as a substitute for $\sigma^z$ on the link qubits. The $B_p$ stabilizers of the surface code defined on link qubits then correspond to operators
\beq
B_{p_\alpha} = \sigma^z_1 \sigma^z_2 \sigma^z_3 \sigma^z_4 \sigma^z_5 \sigma^z_6
\eeq
on the qubits on the honeycomb lattice for $\alpha \in \{x,y\}$.

The substitute for the $\sigma^x$ operators must clearly anticommute with $K_j$, our substitute for $\sigma^z$. It will therefore have the effect of changing the $\epsilon$ occupancy of the link qubit on which it acts. The $A_v$ operator will then do this for all links around the (triangular lattice) vertex on which it acts. One way to achieve this is to use the product of all $x$-links around the vertex. The $A_v$ stabilizers of the surface code then correspond to operators
\beq
A_{p_z} = \sigma^x_1 \sigma^x_2 \sigma^x_3 \sigma^x_4 \sigma^x_5 \sigma^x_6
\eeq
on the qubits of the corresponding $p_z$ plaquette in the honeycomb lattice.

Given the above construction, each plaquette of the honeycomb lattice will have two corresponding stabilizer generators. One is the $W_{p_\alpha}$ of the underlying matching code and the other is the $A_{p_\alpha}$ or $B_{p_\alpha}$ of the embedded surface code. Each are isotropic tensor products of Pauli operators on the six qubits around the hexagonal plaquette, and each correspond to a different Pauli operator. These two Paulis are not the same for every plaquette.

To simplify the description of the code, let us first introduce a third stabilizer operator for each plaquette. This simply corresponds to a product of the other two. Each plaquette then has three stabilizer operators: $(\sigma^x)^{\otimes 6}$, $(\sigma^y)^{\otimes 6}$ and $(\sigma^z)^{\otimes 6}$. Clearly these can also be generated by an alternate pair of stabilizer generators for each plaquette
\beq
S^{\alpha}_p = \sigma^\alpha_1 \sigma^\alpha_2 \sigma^\alpha_3 \sigma^\alpha_4 \sigma^\alpha_5 \sigma^\alpha_6, \,\, \alpha \in \{x,y\}.
\eeq
With these the two generators are the same for every plaquette. These are precisely the stabilizer generators of the well-known color code \cite{bombin_color}.

It is known that the anyon model of the color code is $D(Z_2 \times Z_2)$, which is equivalent to two independent copies of $D({\mathbb Z}_2)$ \cite{bombin_poulin}. However, note that these do not simply correspond to the indepedent anyon models contributed by the matching code and surface code. Instead these two anyon models are combined and reorganized by the embedding procedure. For example, in the original matching code the anyons that live on $x$ and $y$ plaquettes are of the same type, and can be moved from one position to the other. After the embedding, however, the plaquette anyons in these positions will correspond to different types. Any attempt to move an anyon from one position to the other will result in the creation of anyons within the surface code.

It is also important to note that many properties of the color code are not be the same as other codes that realize the $D({\mathbb Z}_2 \times {\mathbb Z}_2)$ anyon model, such as two independent surface codes. One important difference is the wealth of transversal gates possible for color codes \cite{bombin_color}. The technique of embedding codes within others may therefore yield other new codes with advantageous properties.

\section{Conclusions}

Codes that realize the $D({\mathbb Z}_2)$ anyon model are well known, as is the notion that non-Abelian Majorana modes (also known as Ising anyons) may be engineered inside these Abelian codes. However, the means by which the Majoranas are introduced can often be both theoretically and practically inelegant. Often deformations of the lattice are required, making it seem that the Majoranas are a foreign notion introduced to the codes by hand. Also, moving the Majoranas around the code leaves a visible trail along the path they have taken. This obscures their topological nature as non-Abelian Ising anyons, for which paths should be unphysical.

Here we introduce a family of codes in which the Majorana modes occur in a more natural way. The codes can be interpreted as having the Majoranas present at all times, pinned to the vertices of the lattice, without the need for deformations to introduce them. They can be moved around using anyonic state teleportation, allowing their non-Abelian braiding to become evident. The Majoranas will still leave visible trails behind when moved. However, this is simply an artifact of the way braiding is performed by teleportation. The same would occur for any realization of Ising anyons when braiding is performed using this method.

The matching codes should hopefully make it easier to treat these Majorana modes, both theoretically and perhaps even in the lab. For the latter, these codes allow a proof of principle experiment for Majorana braiding to be done with just three qubits and only one and two qubit operations.

Beyond the Majorana interpretation, these codes could also serve another practical use. The potential that $D({\mathbb Z}_2)$ codes have for quantum computation is well recognized, though most focus is on the square lattice planar variant of the surface code. However, it is known that this does not provide the best protection against every error model \cite{beat,abbas,kay_wen}. The matching codes therefore provide a new family of codes to consider for the optimization of error correction against physical error models.

\section*{acknowledgments}

The author would like to thank Benjamin Brown, Tobias Fuhrer and Daniel Loss for discussions that were crucial in shaping this project, and also Konstantinos Meichanetzidis for pointing out a reference. This work was supported by the Swiss National Science Foundation and NCCR QSIT.

\bibliography{refs}

\end{document}